\documentclass{jfm}
\usepackage{graphicx}
\usepackage{epstopdf, epsfig}

\shorttitle{Colloidal sedimentation dynamics in active suspensions}
\shortauthor{B.T. Maldonado, S. Pradeep, R. Ran, D. Jerolmack and P. Arratia}

\title{Sedimentation dynamics of passive particles in dilute bacterial suspensions: emergence of bioconvection}

\author{Bryan O. Torres Maldonado\aff{1}, Shravan Pradeep\aff{2}, Ranjiangshang Ran\aff{1}, Douglas Jerolmack \aff{1,2} \and Paulo E. Arratia\aff{1} \corresp{\email{parratia@seas.upenn.edu}} }

\affiliation{\aff{1}Department of Mechanical Engineering \& Applied Mechanics, University of Pennsylvania, Philadelphia, Pennsylvania 19104, USA
\aff{2}Department of Earth and Environmental Science, University of Pennsylvania, Philadelphia, Pennsylvania 19104, USA}

\DeclareFontFamily{U}{euc}{}
\DeclareFontShape{U}{euc}{m}{n}{<-6>eurm5<6-8>eurm7<8->eurm10}{}%
\DeclareSymbolFont{AMSc}{U}{euc}{m}{n} 
\DeclareMathSymbol{\umu}{\mathord}{AMSc}{"16}

\begin{document}

\maketitle

\begin{abstract}
Microorganisms are ubiquitous in nature and technology. They inhabit diverse environments ranging from small river tributaries and lakes to oceans, as well as wastewater treatment plants and food manufacturing. In many of these environments, microorganisms coexist with settling particles. Here, we investigate the effects of microbial activity (swimming \textit{E. coli}) on the settling dynamics of passive colloidal particles using particle tracking methods. Our results reveal the existence of two distinct regimes in the correlation length scale ($L_u$) and the effective diffusivity of the colloidal particles ($D_{e\!f\!f}$), with increasing bacterial concentration ($\phi_b$). At low $\phi_b$, the parameters $L_u$ and $D_{e\!f\!f}$ monotonically increases with increasing $\phi_b$. Beyond a critical $\phi_b$, second regime is found in which both $D_{e\!f\!f}$ and $L_u$ are independent of $\phi_b$. 
We demonstrate that the transition between these regimes is characterized by the emergence of bioconvection. We use experimentally-measured particle-scale quantities ($L_u$, $D_{e\!f\!f}$) to predict the critical bacterial concentration for the diffusion-bioconvection transition. 

\end{abstract}


\section{Introduction}
Microorganisms are ubiquitous in both natural environments and technological applications. They are highly adaptable and inhabit diverse ecosystems such as swamps, oceans, and rivers [\cite{schallenberg_ecology_1993,zhang_impact_2020,roberto_sediment_2018,nealson_sediment_1997,herndl_microbial_2013}]. They also play a significant role in numerous technological applications [\cite{falkowski_tiny_2012,biofuel_nature,falkowski_biogeochemical_1998}] including fermentation processes for vaccine and food production and wastewater treatment processes [\cite{bacteria_food,bacteria_protein,Water_treatment_science, honda_transition_2023}]. Controlling the sedimentation process in biofuel production, for instance, can lead to improvements in algae separation from the fluid medium and an accumulation of algal biomass at the bottom of the production pond [\cite{biofuel_nature}]. 

Suspensions of swimming microorganisms (e.g., \textit{E. coli}) or active particles (e.g., phoretic colloids) are typical examples of so-called active fluids [\cite{Ramaswamy_2010_ARCM, marchetti_review,PattesonOpinion, moran_2017_artificial_review}].  In these fluids, the constituents inject energy, generate mechanical stresses, and create flows within the fluid medium. Thus, active fluids are systems that are inherently out of equilibrium even in the absence of external forces [\cite{bacteria_non_equilibrium}].  Bacterial activity can lead to many complex phenomena such as reduced shear viscosity [\cite{Gachelin_PRL_2013,lopez_turning_2015,rafai_effective_2010}], hindrance in mixing and transport of passive scalars [\cite{Ran_PNAS_2021}], formation of biofilms [\cite{bacteria_biofilm}], and diffusion enhancement. [\cite{leptos_dynamics_2009,Mino_PRL_2011,Poon_PRE_2013,kurtuldu_enhancement_2011,THIFFEAULT20103487,Kim_POF_2004}]. Particle (or tracer) diffusivity, in particular, has been shown to increase linearly with bacteria concentration (in the dilute regime) \cite{Wu_PRL_2000} but to depend on microorganism swimming behavior \cite{Yodh_bacteriaPRL_2007} and particle size \cite{Patteson2016}. Less understood are the effects of bacterial activity on the settling dynamics of tracer particles.

Initial studies have largely focused on steady-state analysis \cite{palacci_sedimentation_2010,ginot_nonequilibrium_2015,Leonardo2013}. These include sedimentation studies with active Janus particles that show (steady-state) density profiles that decay exponentially with system height
[\cite{palacci_sedimentation_2010,ginot_sedimentation_2018,ginot_nonequilibrium_2015}]. The resulting length scale is larger than expected for thermal equilibrium systems, which is explained through the introduction of an effective temperature (and diffusivity) with values that far exceeds that of passive systems [\cite{palacci_sedimentation_2010}]. These findings are corroborated by theory and simulations with no (or limited) hydrodynamic interactions [\cite{ginot_sedimentation_2018,Cates2010,Cates2008, Tsao2014,vachier_dynamics_2019}]. Experiments with swimming \textit{E. coli} show that bacteria aggregation (due to extra-cellular polymer) can significantly enhanced sedimentation rates  [\cite{Leonardo2013}], while particles settling in suspensions of \textit{C. reinhardtii} show the familiar exponential concentration profile but with an effective gravitational length (or diffusivity) that is proportional to algae concentration \cite{Polin2016}. These studies show that the concept of effective diffusivity (and temperature) can be useful in characterizing the steady sedimentation profiles of active suspensions.

Only recently the time-dependent sedimentation (i.e.,concentration) profiles of bacterial suspensions have been investigated [\cite{Singh_Soft_Matt_2021, Torres_PoF_2022}]. These studies show that, in the dilute regime, bacteria activity can significantly hinder the sedimentation process [\cite{Singh_Soft_Matt_2021}].  At vanishingly low bacteria concentration ($\phi_b$), the (time-dependent) concentration profiles can be accurately described by an advection-diffusion equation; however, as $\phi_b$ is increased, sink-source terms must be introduced in the formulation to account for bacteria population dynamics, including a dispersivity parameter that increases with $\phi_b$ [\cite{Singh_Soft_Matt_2021}]. In this regime, the bacterial suspension undergoes phase separation and bacteria setting speed is found to be strongly correlated to a timescale associated with oxygen depletion within the system [\cite{Torres_PoF_2022}]. These studies highlight the role of oxygen in the sedimentation of living suspensions; microorganisms are known to exhibit aerotaxis, a phenomenon that prompts aerobic bacteria to seek an oxygen source [\cite{oxygen_gravity_bacteria_Wager_1911,Alex_2012_loc,bacteria_aerotaxis_pre_2022,bacteria_aerotaxis_JOP_1959}]. When suspended in a relatively deep layer of fluid, microorganism can deplete the oxygen concentration in the water while oxygen continually diffuses in from the air-water interface above. In response, microorganisms swim towards the oxygen concentration gradient and generate fluid motion, a phenomenon referred to as bioconvection driven by aerotaxis [\cite{bioconvection_bacillus_PRE_1998,hillesdon_development_1995}]. During bioconvection, oxygen-enriched water descends from the interface, giving rise to a complex convection-dependent gradient that guides the swimming bacteria. Analysis predict that the emergence of bioconvection occurs beyond a critical Rayleigh number [\cite{hillesdon_pedley_1996}], a dimensionless number that quantifies the ratio of the buoyancy forces to viscous forces [\cite{hill_pedley_kessler_1989}]. While extensive research has focused on elucidating the mechanisms behind bioconvection formation [\cite{Martin_2020_review_bioconvection,Hill_2005,Pedley_ARFM_1992}], limited experimental work has been conducted to corroborate these theoretical predictions.  


In this contribution, we experimentally investigate the sedimentation dynamics of colloidal particle and the formation of bioconvection patterns in dilute bacterial suspensions using particle tracking methods. We find that bacterial activity has a significant impact on particle diffusivity, leading to the identification of two dynamical regimes. Initially, as bacterial concentration is increased, we observed a corresponding linear increase in diffusivity. However, as bacterial concentration is further increased (but still dilute), particle diffusivity becomes independent of bacterial concentration. This transition coincides with the inception of bioconvection patterns above a critical Rayleigh number ($\Gamma_{cr}$) [\cite{hillesdon_pedley_1996}]. 
 
\section{Materials and Methods}\label{sec:Method}

\begin{figure}
\begin{center}
\centerline{\includegraphics[width=5.3in]{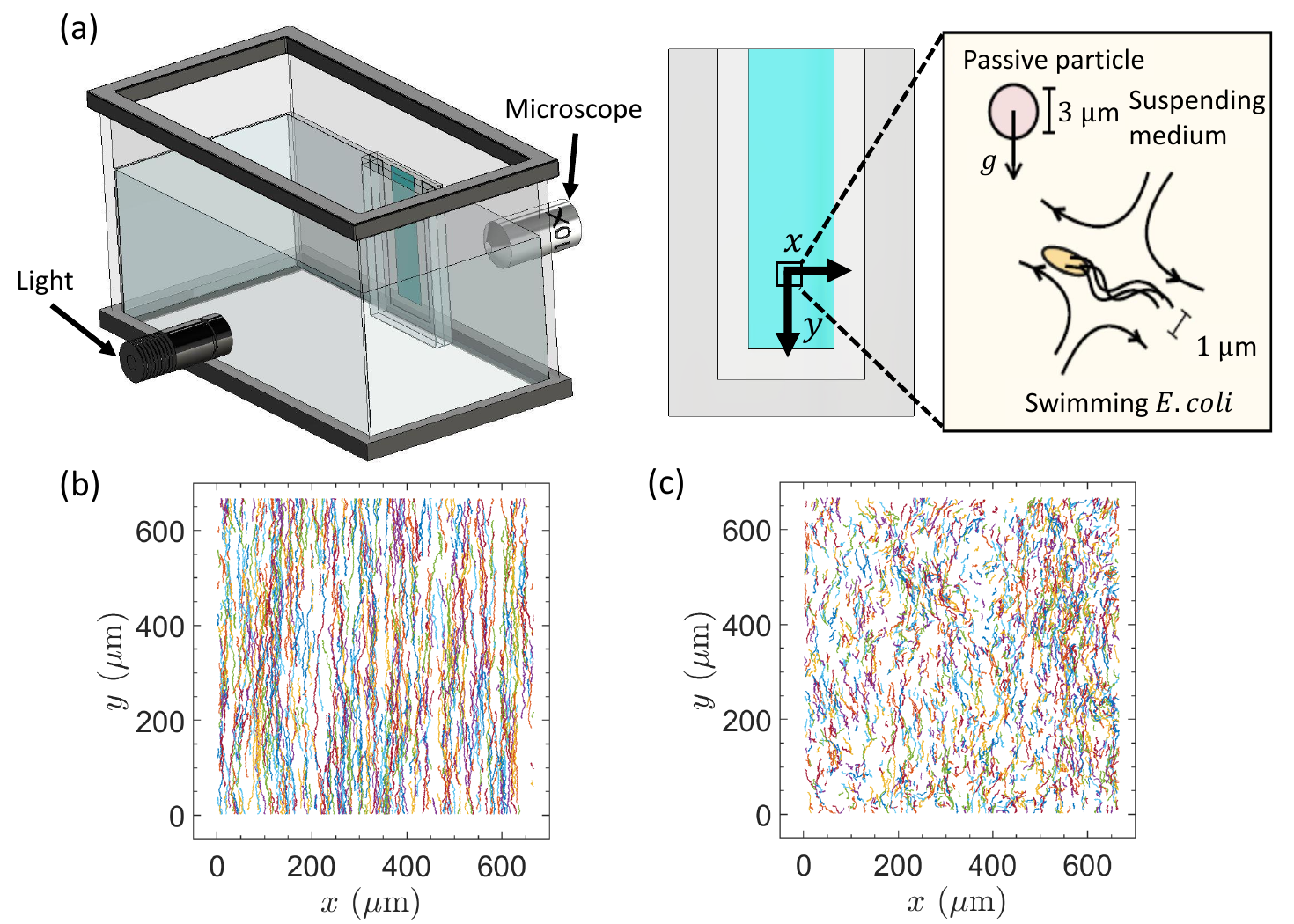}}
 \caption{Experimental setup and sample particle trajectories: (a) A schematic of the setup and bacteria/particle suspensions (active fluid). Sedimentation experiments are conducted in an optically clear square container. The particles are $3.2~\umu$m polystyrene spheres, and the bacteria are $2~\umu$m rod-shaped \textit{E. coli}, both are subjected to gravity. Particles located at the center and at a height $\approx 20~$mm from the bottom of the container are observed through a microscope. These spherical colloidal particles are tracked through particle tracking velocimetry method. Particle trajectories ($\phi_p =0.04\%$) are shown in (b) without bacteria and in (c) with bacteria ($\phi_{b} =0.45\%$). When particles and \textit{E. coli} are combined, the passive particles trajectories in the lateral direction undergo significant modifications compared to their trajectories in the absence of bacteria.
 }
\label{Fig_4_p1}
\end{center}
\end{figure}

We experimentally investigate the sedimentation dynamics of spherical colloids in the presence of bacterial activity. Active fluids are mixtures of bacterial suspensions and passive particles suspended in deionized (DI) water to reach the desired volume fractions of bacteria and particles, denoted as $\phi_b$ and $\phi_p$. Bacterial suspensions are prepared using wild-type \textit{Escherichia coli} (K12 MG1655), which are cultured to saturation ($10^9$ cells/mL) in LB broth (Sigma-Aldrich). Passive particles are polystyrene spheres (density $\rho_p =1.05$~g/cm${}^3$, Thermo Scientific) with a diameter of $d = 3.2$~$\umu$m, which are gently cleaned by centrifugation (Centrifuge 5430, Eppendorf). All active suspensions are contained within a custom-made quasi two-dimensional sedimentation cell constructed from optically transparent polymethyl methacrylate (PMMA) material, offering optimal clarity [see Fig.~\ref{Fig_4_p1}(a)]. The container measures $100~$mm in height, $28~$mm in width, and $400~\umu$m in thickness and is assembled using precision laser cutting and chemical bonding techniques [\cite{sun_rapid_2007}].

Active fluids are manually homogenized using a pipette before being introduced into the custom-made PMMA container. A volume of $\approx1.1$~mL of the prepared fluid is then transferred into the container, as schematically depicted in Fig.~\ref{Fig_4_p1}(a). Subsequently, the container is immersed in a water tank to maintain the fluid at a room temperature of $24$~\textdegree C. Images are captured at a frame rate of $30~$fps with a high-resolution camera (IO Industries, Flare 4M180) and a microscope (Infinity, K2) equipped with a $10\times$ objective. The light source used is a single-color cold visible LED (M490L4, Thor Labs). The initial volume fraction $\phi_b$ of motile bacteria in the container varies from $0\%$ to $0.75\%$, while the volume fraction of spherical colloids is consistently maintained at $\phi_p=0.04\%$. It is worth noting that both $\phi_b$ and $\phi_p$ concentrations are within the dilute regime ($\phi<1\%$), and no significant macroscopic collective behavior is observed in these bacteria-particle suspensions under the microscope \cite{Singh_Soft_Matt_2021,Patteson2016}. The selected range of $\phi_b$ in this study remains below the threshold at which collective motion is typically observed ($\approx 10^{10}$~cells/mL) [\cite{kasyap_hydrodynamic_2014}]. All captured images for the suspensions are acquired within a 30-minute time frame. Importantly, within this designated time frame, no noteworthy reduction in bacteria motility is observed, a finding that has been corroborated by recent investigations [\cite{Ran_PNAS_2021,Singh_Soft_Matt_2021}].


\section{Results and Discussion}\label{sec:Discussion}

The main goal is understand the interplay between bacterial activity and the settling dynamics of passive spherical particles. To this end, we use particle tracking techniques to visualize colloidal particle's Lagrangian trajectories [\cite{crocker_methods_1996}]. Sample particle trajectories from passive ($\phi_p=0.04\%$, no bacteria) and active ($\phi_p=0.04\%$, $\phi_b=0.45\%$) fluids are shown in Fig.~\ref{Fig_4_p1}(b) and (c), respectively. The passive fluid case show particle trajectories predominately in the downward direction ($-y$) with apparent noise along the $x$-axis.  For the active fluid, on the other hand, particle paths show amplified displacement along the $x$-axis, indicating that bacterial activity induces modifications to the colloidal particle downward trajectories. This heightened trajectory enhancement (in the $x$-axis) consistently manifests itself across all experiments, providing an explanation for the previously observed hindrance in the speed of the particle front [\cite{Singh_Soft_Matt_2021,Torres_PoF_2022}]. These qualitative findings show that bacteria modify the settling trajectories of colloidal particles through hydrodynamic interactions, even in the dilute regime ($\phi_b<1\%$). 


\subsection{Characterizing Length Scales of Particle Correlations}
\begin{figure}
\begin{center}
\includegraphics[width=5.3in]{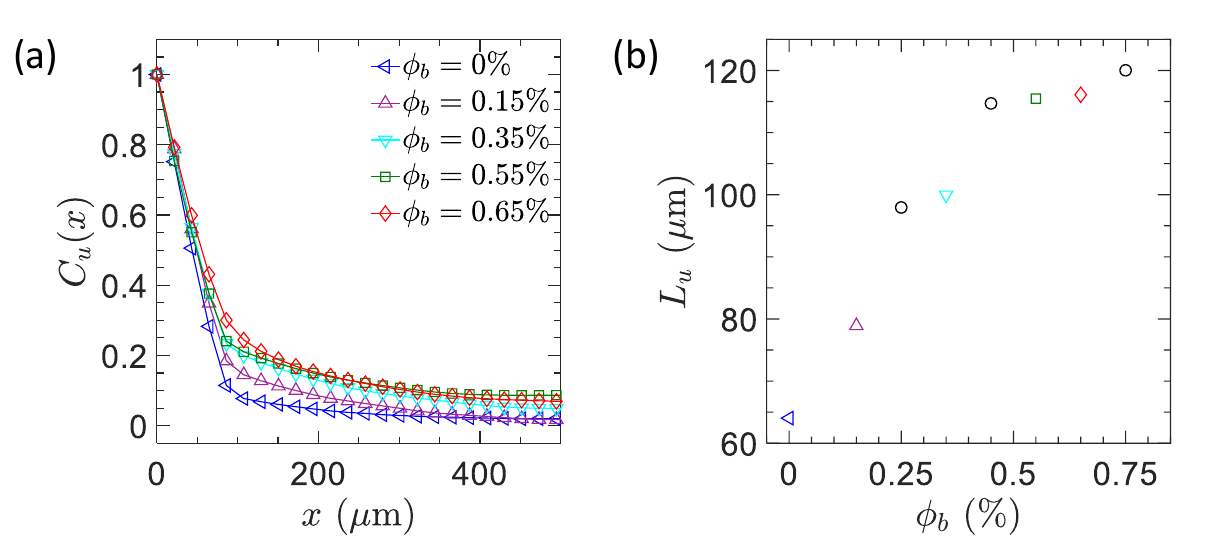}
 \caption{(a) The spatial correlation functions of particle velocities ($u$) in the lateral direction across the $x$-axis. (b) The integral length scale of the lateral velocities at different bacteria volume fraction $\phi_b$. Presence of bacteria leads to an increase in correlation functions, and subsequently, to an increase in length scales.}
\label{Fig_4_p2}
\end{center}
\end{figure}

We quantify the sedimentation dynamics of our fluid by computing their time-dependent velocity fields using an in-house particle imaging velocimetry (PIV) method [\cite{Ran_Polymer_2022,Li_NC_2021,Brosseau_PoF_2022}]. The goal is to understand how bacterial activity ($\phi_b$) alters the spatial correlation functions of particle fluctuations [\cite{segre_1997_PRL,weitz_1997_PRL}]. Here, spatial correlation functions of the velocity fluctuations, $C_u(r_x)$, is defined as:
\begin{equation}
    C_u(r_x)=\langle{u}(x)\,{u}({x}+{r_x})\rangle/\langle u^2\rangle,
\end{equation}
where $x$ represents the direction perpendicular to gravity, $\langle...\rangle$ denotes an ensemble average across all spherical particles, and ${u}({x})$ denotes the local velocity of a particle at position ${x}$. Notably, ${u}({x})$ incorporates contributions from both particle self-diffusivity and hydrodynamic interactions. In the absence of bacteria ($\phi_b=0\%$), we observe a decay in the spatial correlation length scales at small distances (Fig.~\ref{Fig_4_p2}a). With an increase in $\phi_b$, we identify a corresponding increase in spatial correlation functions over larger distances. However, as $\phi_b\geq0.45\%$, the spatial correlation functions becomes independent of $\phi_b$. To quantitatively analyze this trend, we calculate the integral length scale of velocity $L_u$ as shown in Fig.~\ref{Fig_4_p2}(b). This integral length scale is defined as $L_u=\int_0^\infty C_u(r_x)\,dr_x$. Results shows that $L_u$ increases linearly as $\phi_b$ is increased [Fig.~\ref{Fig_4_p2}(b)], but this relationship weakens and $L_u$ becomes independent of $\phi_b$ for $\phi_b\geq0.45\%$. These findings show a positive correlation between $L_u$ and $\phi_b$ up to $\phi_b\geq0.45\%$. This outcome aligns with previous research, highlighting the presence of two distinct regimes within the sedimentation process [\cite{Torres_PoF_2022}].

\subsection{Mean Square Displacements and Diffusivity}
\begin{figure}
\begin{center}
\includegraphics[width=5.3in]{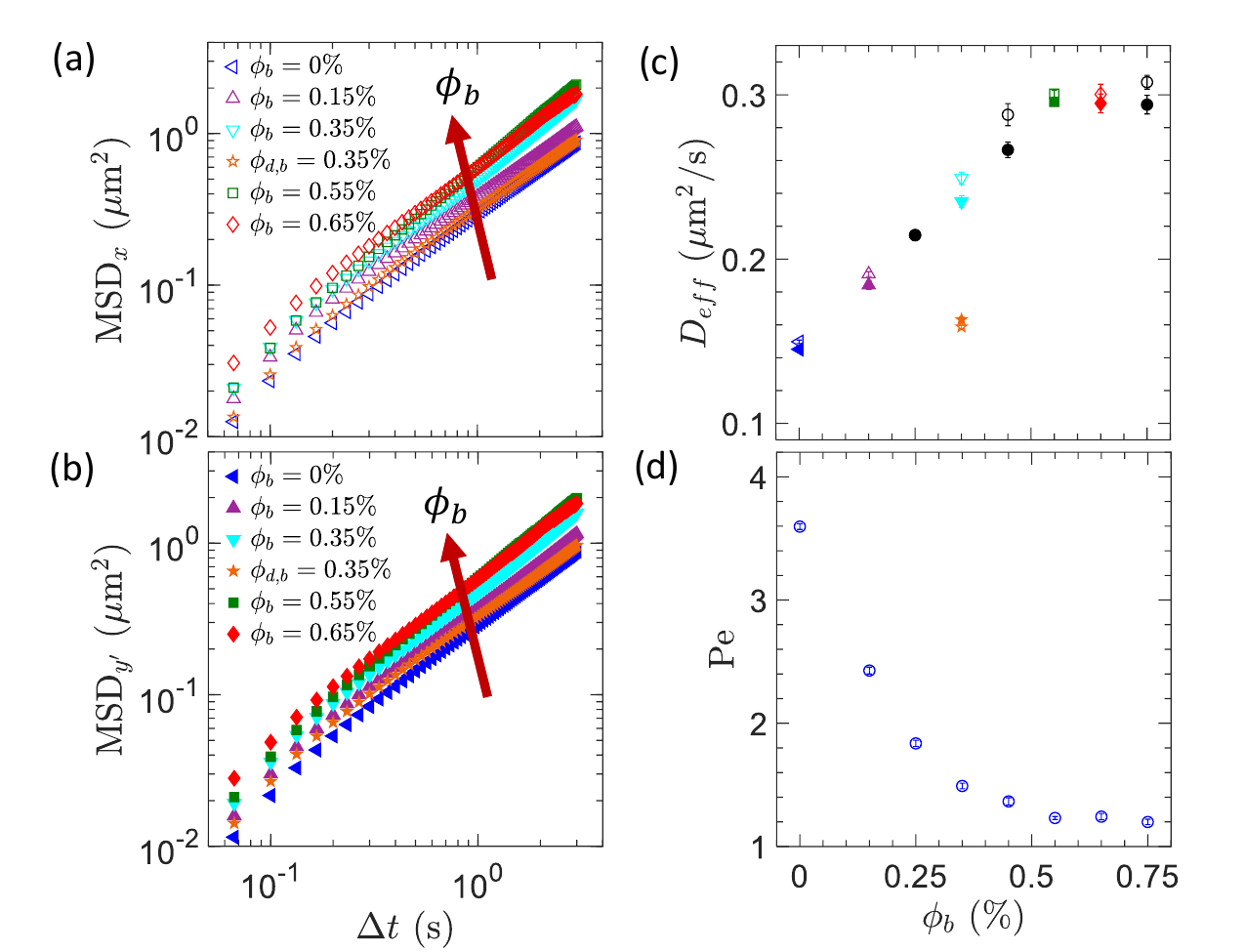}
 \caption{(a) Mean square displacement in the $x$-axis (MSD$_x$) of spherical particles at different bacteria volume fractions $\phi_b$. (b) Mean square displacement of particle fluctuations in the $y$-axis (MSD$_{y^\prime}$) at different $\phi_b$. (c) Effective particle diffusivities $D_{e\!f\!f}$ from MSD$_x$ (Open symbols) and MSD$_y$ (closed symbols) as a function of $\phi_b$. The presence of swimming \textit{E. coli} increases particle fluctuations in both the $x$ and $y$ directions, resulting in higher $D_{e\!f\!f}$. (d) P\'{e}clet number (Pe) as a function of $\phi_b$. The results demonstrate that the presence of bacteria enhances diffusion transport in the settling process, eventually reaching a plateau regime where both advection and diffusion transport are in close balance (Pe$\,\approx 1$).
  }
\label{Fig_4_p3}
\end{center}
\end{figure}

Next, we use experimentally measured particle tracks to compute the mean square displacements (MSD) of the passive particles as a function of $\phi_b$. The mean square displacement is defined as MSD$(\Delta t)$ = $\langle|\textbf{r}(t_R + \Delta t) - \textbf{r}(t_R)|^2\rangle$, where $t_R$ is denoted as the reference time. First, we extract the distance vector \textbf{r} in the direction perpendicular to gravity, $r_x$, yielding the mean square displacement along the $x$-axis (MSD$_x$). As shown in Fig.~\ref{Fig_4_p3}(a), MSD$_x$ exhibits diffusive behavior (at sufficiently long time intervals) that increases with $\phi_b$. We can estimate an effective diffusivity $D_{eff}$ by fitting MSD$_x= 2D_{e\!f\!f}\Delta t$ to the experimental data and find that, similar to $L_u$, $D_{eff}$ grows linearly with $\phi_b$ followed by an asymptote for $\phi_b\geq0.45\%$ (Fig.~\ref{Fig_4_p3}c). Similarly, we compute the mean square displacements of the tracked particles relative to the $y$-axis (MSD$_{y^\prime}$) as a function of $\phi_b$. Notably, we subtract the mean sedimentation rates along the $y$-axis since our focus is on understanding how bacteria activity affect particle fluctuations, i.e., $y^{\prime}=y-\langle y \rangle$. Figure~\ref{Fig_4_p3}(b) shows that MSD$_{y^{\prime}}$ exhibits diffusive behavior that increases with increasing $\phi_b$. We apply a similar fitting approach, relating MSD$_{y^{\prime}}$ to $D_{e\!f\!f}$ as shown in Fig.~\ref{Fig_4_p3}(c) (filled symbols). Our results reveals that for the control case ($\phi_b=0\%$), $D_{e\!f\!f}= 0.150\pm 0.001~\umu$m$^2$/s for MSD$_x$ and $0.145\pm 0.001~\umu$m$^2$/s for MSD$_{y^{\prime}}$. These measurements align with the theoretically predicted value from the Stokes-Einstein relation, $D_0=k_B T/(3 \pi \mu d)=0.150~\umu$m$^2$/s [\cite{Einstein1905}], where $k_B$ represents the Boltzmann constant, $\mu$ is the fluid viscosity, and $T$ denotes the temperature. These results indicate that hydrodynamic interactions among passive particles in our study are relatively weak, which is not surprising given the dilute particle volume fraction ($\phi_p= 0.04\%$).

Overall, the mean square displacement (MSD) measurements of particles in the presence of motile bacteria exhibit an increasing trend as $\phi_b$ rises, as shown in Fig.~\ref{Fig_4_p3}(a) and (b). This enhancement on particle displacement leads to a corresponding increase in $D_{e\!f\!f}$, illustrated in Fig.~\ref{Fig_4_p3}(c). By comparison, a similar experiment was conducted using non-motile bacteria ($\phi_{d,b}=0.35\%$), as shown in Fig.~\ref{Fig_4_p3}(a) and (b). Here, we observe a negligible increase in MSD compared to the absence of bacteria ($\phi_b=0\%$), resulting in $D_{e\!f\!f}= 0.159\pm 0.001~\umu$m$^2$/s along the $x$-axis and $D_{e\!f\!f}= 0.163\pm 0.002~\umu$m$^2$/s along the $y$-axis for $\phi_{d,b}=0.35\%$ [Fig.~\ref{Fig_4_p3}(c)]. This results indicates that the presence of non-motile bacteria exerts negligible impact on particle diffusivity and hydrodynamic interactions. In other words, the observed enhancement in particle diffusivity across the experiments with bacteria is primarily attributed to bacterial activity within the suspension; as mentioned before, $D_{e\!f\!f}$ exhibits an nearly linear relationship with bacteria concentration up to $\phi_b\leq0.45\%$. Surprisingly, however, $D_{e\!f\!f}$ (and $L_u$) become independent of $\phi_b$ for $\phi_b\geq0.45\%$ (Fig.~\ref{Fig_4_p2}b and Fig.~\ref{Fig_4_p3}c). The macroscopic signature of this behavior can be capture by computing the P\'{e}clet number (Pe), defined as Pe$\,=V_p d/(2 \langle D_{e\!f\!f}\rangle)$. Here, $V_p$ is the average settling speed of particles, which has been previously measured for the same fluids and conditions [\cite{Torres_PoF_2022}]. A small Pe signifies the dominance of diffusion in particle transport, while a large Pe indicates the prevalence of advection. Figure~\ref{Fig_4_p3}(d), shows that Pe decreases as $\phi_b$ increases but it reaches a plateau at Pe$\,\approx 1$, indicating a point in which advection and diffusion transport are in close balance. 

This asymptotic behavior has also been observed in macroscopic quantities such as the hindering settling function, $H(\phi)$, for the same type of active fluids and in similar range of $\phi_b$ [\cite{Torres_PoF_2022}]. These results are puzzling since prior studies of (passive) particles in active suspensions demonstrated that $D_{e\!f\!f}$ increases linearly with $\phi_b$ up to $\phi_b\approx1\%$ [\cite{leptos_dynamics_2009,Patteson2016,Wu_PRL_2000}]. A possible explanation is that motile bacteria likely undergo a transition in their swimming motion, shifting from a random pattern to ``organized" trajectories, within the range of $\phi_b\geq0.45\%$ (still dilute). One possibility is the development of bioconvection patterns [\cite{hillesdon_pedley_1996,hill_pedley_kessler_1989}]. We will explore this possibility below.

\subsection{Unveiling the Phenomenon of Bioconvection}
\begin{figure}
\begin{center}
\centerline{\includegraphics[width=5.3in]{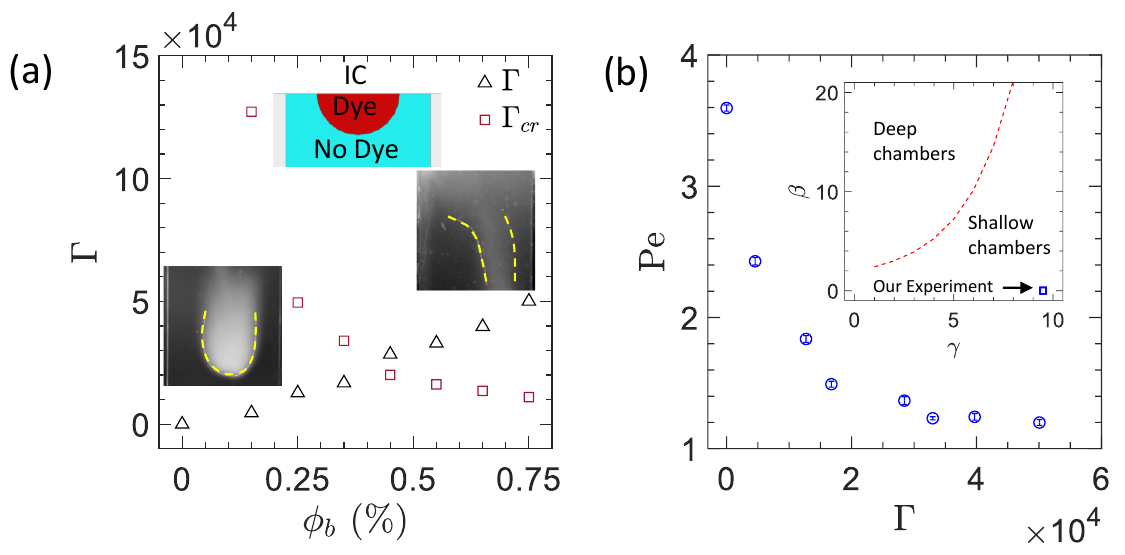}}
 \caption{(a) Experimental Rayleigh number ($\Gamma$), and theoretical critical Rayleigh number ($\Gamma_{cr}$) as a function of bacteria volume fraction ($\phi_b$). Results show a crossover of Ra across $\phi_b\approx0.45\%$. When $\Gamma\,<\,$$\Gamma_{cr}$, swimming bacteria are not significantly affected by the oxygen at the top of the container. However, when $\Gamma\,>\,$$\Gamma_{cr}$, oxygen plays a significant role in the swimming behavior of \textit{E. coli}, leading to bioconvection. This explains the emergence of the plateau regime observed in settling experiments when $\phi_b\geq0.45\%$. (b) P\'{e}clet number (Pe) as a function of experimental Rayleigh number ($\Gamma$), where it shows that convective transport of colloidal particles is reduced with increased bacterial-driven convection. The inset shows the validity of the Eq.~\ref{chap_4_Ra_cr} by using the inequality $\gamma \beta \leq \xi \, $tan$^{-1}\, \xi$.
  }
\label{Fig_4_p4}
\end{center}
\end{figure}

Our study involves a suspension containing \textit{E. coli}, which is known for its oxygen-seeking behavior termed aerotaxis. Following previous work [\cite{hillesdon_pedley_1996}], we define a dimensionless parameter $\Gamma$ that is analogous to the Raleigh number in thermal convection problems:
\begin{equation}
    \mathrm{\Gamma}=\frac{(\rho_b-\rho_w) \phi_b g L_u^3}{\rho_w \nu D_{e\!f\!f} },
    \label{chap_4_Ra}
\end{equation}
where $\rho_b=1.105~$g/cm$^3$ [\cite{DENSITY_ecoli_1981}] and $\rho_w = 1.0$ g/cm$^3$ represents the densities of bacteria and water respectively; $g$ is the constant of gravity, and $\nu$ denotes the kinematic viscosity of water. The only difference between our formulation and prior work is the length scale we employ [\cite{hillesdon_pedley_1996}]. Our present approach incorporates a length scale associated with the particle dynamics (i.e., $L_u$), which is directly measured from the spatial correlation functions. Figure~\ref{Fig_4_p4}(a) show that $\Gamma$ increases linearly with $\phi_b$, as expected. 

Next, following a linear stability analysis, one can define a critical Rayleigh-like number ($\Gamma_{cr}$)  that describes relative importance of bacterial diffusion to bioconvection (aerotatic behavior plus density difference) behavior [\cite{hillesdon_pedley_1996}]. Initially, it is necessary to establish two dimensionless constants stemming from the equations governing cell and oxygen conservation. The first of these constants, denoted as $\beta$, characterizes the relationship between the rate of oxygen consumption and the rate of oxygen diffusion. It is defined as:
\begin{equation}
    \beta=\frac{K_0 n_0 L_u^2} {D_{c} C_0},
        \label{chap_4_beta}
\end{equation}
where $K_0$ is the oxygen consumption rate, $n_0$ is the initial cell concentration, $D_{c}$ represents oxygen diffusivity, and $C_0$ denotes the initial oxygen concentration. Here, we assume $K_0\approx2\times10^{-18}~$mol/min/cell [\cite{schwarz-linek_escherichia_2016}], $D_{c}\approx200~\umu$m$^2$/s [\cite{David_1996_oxygen_diff}], and $C_0\approx0.0058~$mLO$_2/$mLH$_2$O [\cite{Carpenter1966NEWMO}].

The second dimensionless constant, $\gamma$, quantifies the strength of oxytactic swimming relative to random diffusive swimming. It is expressed as:
\begin{equation}
    \gamma=\frac{cV_{sb}} {D_b},
    \label{chap_4_gamma}
\end{equation}
where $c$ is the chemotaxis constant, $V_{sb}$ represents bacteria swimming speed, and $D_{b}$ is the bacteria diffusivity. Here, we assume $c=V_{sb}\tau_r$, where $\tau_r=0.95~$s stands for the mean run time of \textit{E. coli} [\cite{PattesonSciRep}]. Additionally, we take $D_{b}\approx10~\umu$m$^2$/s and $V_{sb}\approx10~ \umu$m/s [\cite{Ran_PNAS_2021}].

\begin{figure}[t!]
\begin{center}
\includegraphics[width=4in]{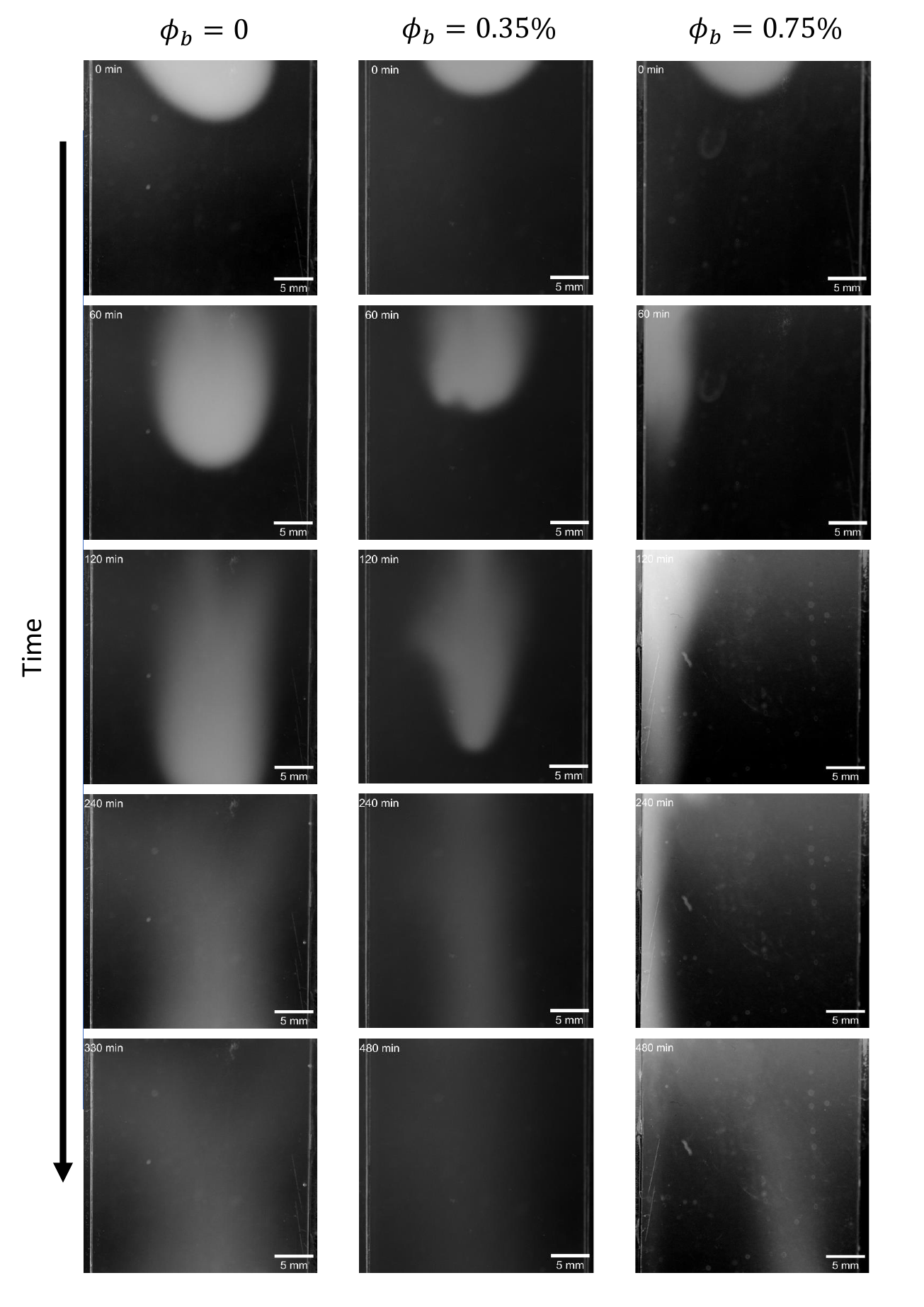}
 \caption{Representative snapshots of the time evolution of the settling suspensions with dye at the top of the settling container. Left column: No bacteria ($\phi_b=0\%$); middle column: active bacteria at $\phi_b=0.35\%$; right column: active bacteria at $\phi_b=0.75\%$. Dye in passive suspensions shows that it settles straight downwards as a function of time. When bacteria are added, as shown in the middle column ($\phi_b<0.45\%$), the results show that dye settling is slightly hindered due to the random motion of bacteria. However, no significant bioconvection patterns were observed. When $\phi_b\geq0.45\%$, as shown in the right column, dye gets convected in a counterclockwise motion. Dye return to the field of view from below at long times, showing that bioconvection is observed in this case.}
\label{Fig_4_p5}
\end{center}
\end{figure}

Our experiment can be approached as a shallow chamber, since it satisfy the inequality $\gamma \beta \leq \xi \, $tan$^{-1}\, \xi$, wherein $\xi$ is defined as $\xi^2= e^\gamma -1$ [Fig.~\ref{Fig_4_p4}(b), inset]. Therefore, the analytical approximation, following [\cite{hillesdon_pedley_1996}], $\Gamma_{cr}$ can be expressed as:
\begin{equation}
    {\Gamma_{cr}}=\frac{576}{\gamma \beta }.
     \label{chap_4_Ra_cr}
\end{equation}
We proceed to compute the experimental values of $\Gamma$ and  $\Gamma_{cr}$ as a function of $\phi_b$, as shown in Fig.~\ref{Fig_4_p5}(a). The data illustrates an increase in $\Gamma$ with rising $\phi_b$, while $\Gamma_{cr}$ decreases with higher $\phi_b$. This aligns with the expectation that higher $\phi_b$ leads to accelerated oxygen consumption by \textit{E. coli} within the fluid.  Remarkably, an intersection between $\Gamma$ and $\Gamma_{cr}$ occurs around $\phi_b\approx0.45\%$, signifying the onset of bioconvection when $\phi_b\geq0.45\%$. This observation could elucidate the plateau observed in $D_{e\!f\!f}$ and $L_u$, and in the macroscopic hindering settling function ($H(\phi)$) seen in previous experiments [\cite{Torres_PoF_2022}]. Concurrently, with the increase in $\Gamma$, the P\'{e}clet number (Pe) reaches a plateau at unity for $\phi_b\geq0.45\%$, as shown in Fig.~\ref{Fig_4_p5}(b). This strongly suggests that the reduction in convective transport of colloidal particles (Pe) results from the increased convective motion driven by swimming bacteria.

To further illustrate the existence of bioconvection in settling active fluids, we perform dye mixing experiments for a range of $\phi_b$. These experiments involved introducing $100~\umu$L of dye ($2.5\times10^{-3}M$ fluorescein aqueous solution) into bacterial solutions at the upper portion of the square chamber within the experimental setup. Image capture occurred every 2 minutes over a 14-hour period using a Nikon D7100 camera equipped with a 105~mm Sigma lens. Illumination was achieved using black light (USHIO, F8T5/BLB), peaking at $368~$nm in the ultraviolet (UV) range. Importantly, $90\%$ of the UV energy falls within the long-wave UVA-I range (340 to 400 nm), and it minimally impacted suspension activity [\cite{Ran_PNAS_2021,Vermeulen_2008}]. As expected, experiments with no bacteria ($\phi_b=0\%$) demonstrated dye settling dominated by diffusion and moving directly downward, evident in the left column of Fig.~\ref{Fig_4_p5}. Similarly, $\phi_b=0.35\%$ experiments, with $\Gamma ~<~ \Gamma_{cr}$, displayed no significant convection patterns (middle column). However, experiments at $\phi_b=0.75\%$, where $\Gamma ~>~ \Gamma_{cr}$, exhibited pronounced convection patterns (right column). In the early stages, the dye exhibited counterclockwise motion due to convection, transitioning to complex bioconvection patterns in the upward direction at later times. 

\begin{figure}
\begin{center}
\includegraphics[width=3in]{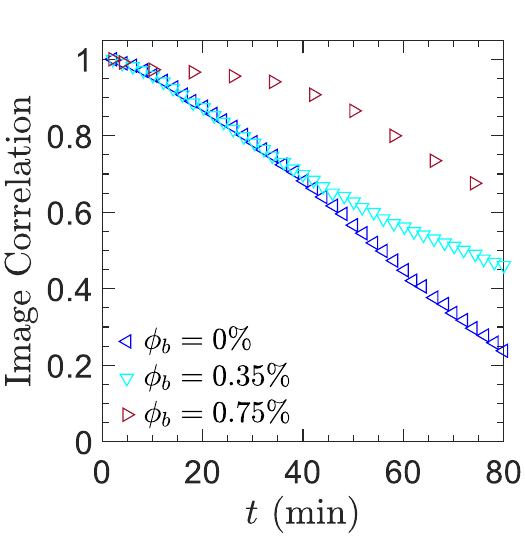}
 \caption{Temporal evolution of image correlation in dye experiments as a function of time, for no bacteria ($\phi_b=0\%$), $\phi_b=0.35\%$, and $\phi_b=0.75\%$. For concentrations $\phi_b<0.45\%$, the correlation is reminiscent of pure diffusive behavior throughout most of the observation period. However, for $\phi_b>0.45\%$, the image correlation showed a significant delay due to bacteria, which convects the dye perpendicular to gravity's direction and subsequently exhibits diffusive behavior.
  }
\label{Fig_4_p6}
\end{center}
\end{figure}

We quantify the dynamics observed in these images by computing the image correlation with respect to the initial image ($t=0$ min). Figure~\ref{Fig_4_p6} illustrates that in the absence of bacteria ($\phi_b=0\%$), the correlation is reminiscent of pure diffusive behavior throughout most of the observation period (the time window is selected so that most, if not all, of the initial dye is still in the field of view). For $\phi_b=0.35\%$, decorrelation rates are initially similar to the no-bacteria case but start to deviate and slow down at about $\Delta t \approx$ 30 minutes. That is, bacterial activity hinders the settling behavior of dye particles as the suspension transitions from a diffusive-dominated behavior to one dominated by bacterial-activity induced convection. Macroscopically, this behavior manifests itself by decreasing the hindered settling function $H(\phi)$ (where $H(\phi)=V_p(\phi)/V_0$, $V_p(\phi)$ is the mean sedimentation speed of the particles in the presence of bacteria, and $V_0$ is the Stokes' settling speed) as reported in our earlier work [\cite{Torres_PoF_2022}]. As $\phi_b$ is further increase ($\phi_b=0.75\%$), however,  the correlation data shows two distinct parts: at short times the correlation shows a a significant delay due to bioconvection-induced motion of the dye particles perpendicular to the gravity. At longer time scales image correlations mirror the samples with diffusion-like behavior emerges due to the dye particles following the downward draft in bioconvection rolls. These results shows delicate interplay between bacterial activity, oxygen diffusion (i.e., aerotactic behavior), and density differences. Our results clarifies the existence of two regimes in settling dynamics parameters observed in our present study ($L_u$, $D_p$, and $Pe$) and the previous work ($H(\phi$)) [\cite{Torres_PoF_2022}], indicating that active suspensions are purely diffusive at $\phi_b<0.45\%$ and are characterized with complex bioconvection patterns for $\phi_b\geq0.45\%$.


Our experimental findings demonstrate that the presence of bacteria leads to an increase in both particle correlation length-scales (Fig.~\ref{Fig_4_p2}) and their effective diffusivity values, $D_{e\!f\!f}$ (Fig.~\ref{Fig_4_p3}). As mentioned before, these measurements exhibit a linear increase up to approximately $\phi_b\approx0.45\%$. Beyond this threshold ($\phi_b\geq 0.45\%$), both $L_u$ and $D_{e\!f\!f}$ remain constant, establishing their independence from $\phi_b$. These experimental results align with our prior findings, wherein the hindered settling function $H(\phi)$ of the particle front showed a linear decrease with bacterial concentration for $\phi_b<0.45\%$. The hindered settling function $H(\phi)$ of the particle front ultimately reaches a plateau at $\phi_b\geq 0.45\%$ [\cite{Torres_PoF_2022}]. We assessed the impacts of advection and diffusion by quantifying the P\'{e}clet (Pe) number. Within the linear regime ($\phi_b<0.45\%$), we noticed a decrease in advection effects due to particle hindrance, accompanied by an increase in diffusion effects owing to increased particle fluctuations. In the vicinity of the plateau regime, both settling (advection) and random motion (diffusion) exhibit near-equivalence ($Pe\approx 1$). The emergence of this plateau regime can be explained by Eq.~\ref{chap_4_Ra}, quantifying $\Gamma$, and Eq.~\ref{chap_4_Ra_cr}, determining $\Gamma_{cr}$. Interestingly, our results manifest a crossover around $\phi_b\approx0.45\%$ between the experimental Rayleigh number and the estimated critical Rayleigh number (Fig.~\ref{Fig_4_p4}). This suggests that under $\phi_b<0.45\%$, bacteria engage in random swimming, with minimal oxygen-related effects at the upper portion of the container. However, at $\phi_b\geq0.45\%$, oxygen plays a prominent role, giving rise to intricate bioconvection patterns as depicted in Fig.~\ref{Fig_4_p5}. In both these regimes, the input of energy via bacterial swimming motions significantly influences the sedimentation process, thereby enhancing particle diffusivities.


\section{Conclusion}
We experimentally investigate the sedimentation dynamics of colloidal particles within dilute suspensions of swimming bacteria. Our overall observation reveals that bacterial activity enhances the effective diffusivities ($D_{e\!f\!f}$) and the correlation length $L_u$ of colloidal particles. Moreover, our experimental analysis of spherical particle behavior exhibits two distinct regimes. The first regime showcases a linear increase in both $D_{e\!f\!f}$ and $L_u$ in correlation with bacterial concentration ($\phi_b$) up to $\phi_b<0.45\%$. In contrast, once $\phi_b\geq0.45\%$, both $D_{e\!f\!f}$ and $L_u$ remain unaltered despite varying bacterial concentration. To uncover the foundation of these distinct regimes, we computed the experimental Rayleigh number ($\Gamma$) and the critical Rayleigh number ($\Gamma_{cr}$), respectively, using experimentally measured quantities. Solving for these non-dimensional quantities reveals a transition point in bacterial concentration that divides the two observed regimes within our experiments. This transition suggests that the system becomes unstable beyond $\phi_b\geq0.45\%$, consequently giving rise to complex bioconvection patterns. This inference gains further validation through a separate set of experiments utilizing dye as a tracer within the fluid. For $\phi_b<0.45\%$, the variance demonstrates diffusive behavior. Intriguingly, when $\phi_b\geq0.45\%$, the variance initially exhibits diffusive-like behavior for short times, followed by a transition to convection-like behavior. In summary, our experimental findings shows that the reduction in convective transport of colloidal particles is attributed to increased bacterial-driven convective motions. Future work involving diverse colloidal sizes and swimmer actuation modes (such as pusher vs puller) holds the potential to elucidate whether the insights obtained from our particle-bacteria interaction study can be generalized to others.


\bibliographystyle{jfm}
\bibliography{jfm-instructions}

\end{document}